\documentclass[noinfoline, aos,11pt, reqno]{imsart} 

\RequirePackage[OT1]{fontenc}
\RequirePackage{amsthm,amsmath,natbib}
\usepackage{color,soul}
\RequirePackage[draft=true,colorlinks,linkcolor=blue,urlcolor=blue,citecolor=blue]{hyperref}
\RequirePackage{hypernat}
\usepackage{graphicx}
\usepackage{amsmath,amssymb, amsthm,epsfig,bm}
\usepackage{epstopdf}
\sethlcolor{yellow}
\startlocaldefs
\theoremstyle{plain}
\newtheorem{theorem}{Theorem}

\newtheorem{proposition}[theorem]{Proposition}

\theoremstyle{definition}

\endlocaldefs

\newcommand{\toinL}{\mathop{\to}\limits^{L}}      
\newcommand{\toinP}{\mathop{\to}\limits^{P}}      
\newcommand{\dnorm}{\mathcal{N}}
\newcommand{\sign}{\mbox{ sign}}

\begin{document}
\begin{frontmatter}
\title{On Weight Matrix and Free Energy Models for Sequence Motif Detection\protect\thanksref{T1}}
\runtitle{On Motif Detection}
\thankstext{T1}{To appear in {\em Journal of Computational Biology}.}

\begin{aug}
\author{\fnms{Qing} \snm{Zhou}\thanksref{t1}} 

\thankstext{t1}{Department of Statistics, 8125 Mathematical Sciences Building,
University of California, Los Angeles, CA 90095
(email: zhou@stat.ucla.edu).}
\runauthor{Q. Zhou}

\affiliation{University of California, Los Angeles}

\end{aug}

\begin{abstract}

The problem of motif detection can be formulated as the construction of a discriminant function
to separate sequences of a specific pattern from background. In computational biology, 
motif detection is used to predict DNA binding sites of a transcription factor (TF),  
mostly based on the weight matrix (WM) model or the Gibbs free energy (FE) model.
However, despite the wide applications, theoretical analysis of these two models and their predictions
is still lacking.
We derive asymptotic error rates of prediction procedures based on these models
under different data generation assumptions. 
This allows a theoretical comparison between the WM-based and the FE-based predictions in
terms of asymptotic efficiency. 
Applications of the theoretical results 
are demonstrated with empirical studies on ChIP-seq data and protein binding microarray data.
We find that, irrespective of underlying data generation mechanisms, 
the FE approach shows higher or comparable predictive power relative to the WM approach
when the number of observed binding sites used for  constructing a discriminant decision is not too small.


\end{abstract}

\end{frontmatter}

\section{Introduction}

Transcription factors (TFs), a class of proteins, regulate gene transcription through 
their physical interactions with particular DNA sites. 
Such a DNA site is called a transcription factor binding site
(TFBS), which is usually a short piece of nucleotide sequence (e.g., `CATTGTC'). 
Typically, a TF can bind different sites and regulate a set of genes. 
A key observation is that sites of the same TF share similarity in their sequence composition,
which is characterized by a motif.
Since gene regulation has always been an important
problem in biology, many computational methods have been developed to
predict whether a given DNA sequence can be bound by a TF.
Please see Elnitski et al. (2006), Ji and Wong (2006), and Vingron et al. (2009) for recent reviews on relevant methods.

The prediction of TFBS's considered in this article is formulated as a classification problem.
Denote by $w$ the width of the binding sites and
code the four nucleotide bases, A, C, G and T, by a set of positive integers 
$\mathcal{I}=\{1,\cdots,J\}$ ($J=4$). 
Suppose that we have observed a sample of labeled sequences of length $w$, 
$\bm{D}_n=\{(Y_k,\bm{X}_k)\}_{k=1}^n$,
where $\bm{X}_k \in \mathcal{I}^w$ and $Y_k\in\{0,1\}$ indicating whether
$\bm{X}_k$ is bound by the TF ($Y_k=1$) or not ($Y_k=0$).
We call $\bm{D}^+_n=\{\bm{X}_k : Y_k=1\}$ observed binding sites (or motif sites)
and $\bm{D}^-_n=\{\bm{X}_k : Y_k=0\}$ background sites (or background sequences).
Then, motif detection is to construct a discriminant function from
$\bm{D}_n$ to predict the label of any new sequence $\bm{x} \in \mathcal{I}^w$. 

Most of the existing computational methods for motif detection can be
classified into two groups.  The starting point of the first group is the sequence specificity
of binding sites, which is often summarized by the position-specific
weight matrix (WM). For early developments of WM, please see Stormo (2000).
Under the WM model, each nucleotide (letter) in a binding site is assumed to be generated
independently from a multinomial distribution on $\{\mbox{A, C, G, T}\}$. 
This model has been widely used in search of TFBS's (e.g., Hertz and Stormo, 1999; Kel et al., 2003; 
Rahmann et al., 2003; Turatsinze et al., 2008), {\it de novo}
motif finding (e.g., Stormo and Hartzell, 1989; Lawrence et al., 1993; Bailey and Elkan, 1994; 
Roth et al., 1998; Liu et al., 2002)
and many other works reviewed in Vingron et al. (2009).
The second group aims 
at modeling physical binding affinity between a TF and its binding sites
via the concept of the Gibbs free energy (FE) or binding energy 
(e.g., Berg and von Hippel, 1987; Stormo and Fields, 1998; Gerland et al., 2002; Kinney et al., 2007).
Assuming that each nucleotide in a DNA sequence of length $w$ ($w$-mer) contributes additively
to the interaction with the TF, this approach often leads to a regression-type model
for the conditional distribution of binding affinity given a piece of nucleotide sequence
(e.g., Djordjevic et al. 2003; Foat et al. 2006). This group of methods have 
tight connections with predictive modeling approaches to
gene regulation, reviewed in Bussemaker et al. (2007), which
can be regarded as a natural generalization to the free energy 
framework (Zhou and Liu, 2008). Although the standpoints are different, the two groups of approaches
are in some sense closely related. They often give similar discriminant functions for predicting
TFBS's, and there are many FE-based methods that use a weight matrix to approximate Gibbs free energy
(e.g., Granek and Clarke, 2005; Roider et al., 2007). 

In spite of the fast methodological development on the WM and the FE models,
there is still a lack of solid theoretical analysis to compare model
assumptions, parameter estimations and response predictions of the two approaches. 
Such theoretical analysis can provide insights into these methods by seeking answers
to a series of questions. For example, what are the common and distinct assumptions
between the WM and the FE models, what is the relative performance between the two approaches
in predicting TFBS's given a certain data generation mechanism, and how to calculate
their predictive error rates when the size of observed sample $\bm{D}_n$ becomes large?
Without answering these questions, one may find it difficult to understand
the nature of these methods and cannot extract the full information
contained in extensive empirical comparisons between the two approaches.

In this article, we compare model assumptions and parameter estimations
of typical WM and FE approaches, derive asymptotic
error rates of their predictions under different data generation models,
and perform comparative studies on large-scale binding data.
The article is organized as follows. 
In Section~\ref{sec:model} we review the basic models of the two approaches.
Asymptotic error rates of prediction procedures based on 
these models are derived and analyzed in Section~\ref{sec:prediction}.
Computational approaches are developed in Section~\ref{sec:computation} for practical applications
of the theoretical results.
Numerical analysis and biological applications are presented in Sections~\ref{sec:numericalWMM} and \ref{sec:applications}, 
respectively, with a comparison of the WM-based and the FE-based predictions on ChIP-seq data and protein binding microarray data.
The paper concludes with discussions in Section~\ref{sec:discuss}. Some mathematical details are provided in Appendices.
Although presented in the specific context of motif detection, the results in this article
are generally applicable to the modeling and classification of categorical data. 

\section{Models} \label{sec:model}

Let $c$ be a scalar, $\bm{u}=(u_1,\cdots,u_J)$ be a (column) vector, $\bm{v}=(v_1,\cdots,v_w) \in \mathcal{I}^w$, 
and $\bm{A}=(a_{ij})_{w\times J}$ and $\bm{B}=(b_{ij})_{w\times J}$ be two $w \times J$ matrices. 
For notational ease,
we define $c\pm\bm{A}:= (c\pm a_{ij})_{w\times J}$, 
$\bm{A}/\bm{B}:=(a_{ij}/b_{ij})_{w \times J}$ provided that $b_{ij} \neq 0$,
$\bm{v} \bm{A}:= \sum_{i=1}^w a_{iv_i}$, 
$\bm{A}(\bm{v}) := \prod_{i=1}^w a_{iv_i}$ and $\bm{u}(\bm{v}):=\prod_{i=1}^w u_{v_i}$.
Furthermore, we define $\bm{v}_{[-k]} := (v_1,\cdots,v_{k-1},v_{k+1},\cdots,v_w)$
and $\bm{A}_{[-k]}$ by removing the $k$th row from $\bm{A}$, for $k=1,\cdots,w$.
Symbols `$\toinL$' and `$\toinP$' are used for convergence in law and in probability, respectively.

Let $\bm{\theta}_0=({\theta}_{01},\cdots,\theta_{0J})$ be
the cell probabilities (probability vector) of a multinomial distribution for i.i.d. 
background nucleotides, where
$\sum_{j=1}^J \theta_{0j}=1$ and $\theta_{0j}> 0$ for $j=1,\cdots,J$. Since $\bm{\theta}_0$
can be accurately estimated from a large number of genomic background sequences,
we assume that it is given in the following analyses. 
Throughout the paper, we assume that the cell probabilities of 
any multinomial distribution are bounded away from 0.

\subsection{The weight matrix model} \label{sec:WMM}

Let $\bm{X}=(X_1,\cdots,X_w) \in \mathcal{I}^w$ be a sequence of length $w$.
In the weight matrix model (WMM), we assume that $\bm{X}$ is
generated from a mixture distribution. Let $Y \in \{0,1\}$ label
the mixture component. With probability $q_0$, $Y=0$ and $\bm{X}$
is generated from an i.i.d. background model (with parameter) $\bm{\theta}_0$,
that is, $P(\bm{X} \mid Y=0) = \bm{\theta}_0(\bm{X})$.
With probability $q_1=1-q_0$, $Y=1$ and $\bm{X}$
is generated from 
a weight matrix $\bm{\Theta}=(\theta_{ij})_{w\times J}=(\bm{\theta}_1,\cdots,\bm{\theta}_w)^t$, 
where $\bm{\theta}_i=(\theta_{i1},\cdots,\theta_{iJ})$ is a probability vector for $i=1,\cdots,w$ and
$X_i$ is independent of other $X_k\;(k\neq i)$. To be specific, 
$P(\bm{X} \mid Y=1) = \bm{\Theta}(\bm{X})$.
From this model the log-odds ratio of $Y$ given $\bm{X}$ is
\begin{equation} \label{eq:logpostodds}
\log \frac{P(Y=1\mid \bm{X})}{P(Y=0\mid \bm{X})} 
=\log \frac{q_1\bm{\Theta}(\bm{X})}{q_0 \bm{\theta}_0(\bm{X})}
= \log (q_1/q_0) + \sum_{i=1}^w \log(\theta_{iX_i}/\theta_{0X_i}).
\end{equation}
In the WM-based prediction, $q_1$ is typically fixed by prior expectation
or determined by the relative cost of the two types of errors (false positive vs false negative). 
Effectively, we assume that $q_1$ is given. Let
\begin{equation}  \label{eq:definebeta}
\beta_0 =  \log(q_1/q_0), \;\;
\beta_{ij} = \log(\theta_{ij}/\theta_{0j}),
\end{equation}
for $1\leq i \leq w, 1\leq j \leq J$ and $\bm{\beta}=(\beta_{ij})_{w \times J}$.
We rewrite (\ref{eq:logpostodds}) as
\begin{equation} \label{eq:logpostoddsbybeta}
\log \frac{P(Y=1\mid \bm{X})}{P(Y=0\mid \bm{X})} = \beta_0 + \sum_{i=1}^w \beta_{iX_i} =\beta_0+\bm{X}\bm{\beta}:= h(\bm{X}),
\end{equation}
which defines an additive discriminant function to predict $Y$ given $\bm{X}$, i.e., to predict 
whether the sequence $\bm{X}$ can be bound by the TF. 
The label $Y$ will be predicted as 1 if $h(\bm{X})\geq 0$ and 0 otherwise.
This prediction can be regarded as a naive Bayesian classifier. 

Given observed binding sites $\bm{D}^+_n$, 
we estimate $\bm{\Theta}$ by the maximum likelihood estimator (MLE)
$\hat{\bm{\Theta}}^m=(\hat{\bm{\theta}}^m_1,\cdots,\hat{\bm{\theta}}^m_w)^t$ and substitute it in
equation (\ref{eq:definebeta}) to obtain $\hat{\bm{\beta}}^m$.
Here, the superscript `$m$' stands for estimators based on the WMM.
Let $d \hat{\bm{\theta}}^m_i = \hat{\bm{\theta}}^m_i -  \bm{\theta}_i$, which is an infinitesimal   
in the order of $1/\sqrt{n}$ as $n\to\infty$. The standard asymptotic theory (e.g., Ferguson 1996) implies that %
\begin{equation} \label{eq:dthetamdistrn}
\sqrt{n} d \hat{\bm{\theta}}^m_i  \toinL  \dnorm(\bm{0}, \bm{\Sigma}^m_i ) \mbox{ restricted to } 
		\sum_{j=1}^J d \hat{\theta}^m_{ij} =0, \mbox{ as  $n\rightarrow \infty$},
\end{equation}
and that $\sqrt{n} d \hat{\bm{\theta}}^m_i$, $i=1,\cdots,w$, are mutually independent. 
The $(j,k)$th element of the covariance matrix $\bm{\Sigma}^m_i$ 
is $(\delta_{jk}\theta_{ij}-\theta_{ij}\theta_{ik})/q_1$
where $\delta_{jk}$ is the Kronecker delta symbol and $1\leq j,k \leq J$.
From equation (\ref{eq:definebeta})  we have $d \hat{\beta}^m_{ij} = d \hat{\theta}^m_{ij}/\theta_{ij}$,
which leads to the following limiting distribution,
\begin{equation} \label{eq:limitingbetam}
\sqrt{n} d\hat{\beta}^m_{ij}  \toinL  \dnorm (0, (1-\theta_{ij})/(\theta_{ij}q_1)),\mbox{ for  $j=1,\cdots,J$, as $n\rightarrow \infty$},
\end{equation}
with $\sqrt{n}d\hat{\bm{\beta}}^m_{i}$ mutually independent for $i=1,\cdots, w$.

\subsection{The free energy model} \label{sec:FEM}

Let $F$, $\bm{X}=(X_1,\cdots,X_w)$ and $F\bm{X}$ be a TF, a DNA sequence, 
and the corresponding TF-DNA complex, respectively.
The process of the TF-DNA interaction can be described by the chemical reaction
$F+ \bm{X} = F\bm{X}$.
The concentrations of the three molecules at chemical equilibrium, 
$[F],[\bm{X}]$ and $[F\bm{X}]$, are determined by the association 
constant $K_a(\bm{X})$, that is,
\begin{equation*}
\frac{[F\bm{X}]}{[F] [\bm{X}]}= K_a(\bm{X}) =\exp\left\{ - \frac{\Delta G(\bm{X})}{RT} \right\},
\end{equation*}
where $\Delta G(\bm{X})$ is the Gibbs free energy (FE) for the interaction of $F$ with $\bm{X}$, $R$ is
the gas constant and $T$ the temperature. We regard $RT>0$ as a constant.
Suppose that the contribution of a single nucleotide $X_i$ to the FE 
is additive (von Hippel and Berg, 1986; Benos et al., 2002) so that 
we may write $-\Delta G(\bm{X}) / (RT) = c+\sum_{i=1}^w b_{iX_i}$. Then we have
\begin{equation}\label{eq:freeenergydef}
\log \frac{[F\bm{X}]}{[\bm{X}]} = \log [F] + c + \sum_{i=1}^w b_{iX_i} := b_0 + \sum_{i=1}^w b_{iX_i}.
\end{equation}
To avoid non-identifiability in estimation, we take 
$\bm{S}_{ref}=(s_1, \cdots, s_w)$ as a reference sequence 
to determine a baseline level of the FE, and define
$\tilde{\beta}_{ij} =  b_{ij}-b_{is_i}$ for all $i, j$ and 
$\tilde{\beta}_0 = b_0 + \sum_{i=1}^w b_{is_i}$, such that $\tilde{\beta}_{is_i}\equiv0$ for $i=1,\cdots,w$ and
\begin{equation}\label{eq:reftrans}
b_0 + \sum_{i=1}^w b_{iX_i} = \tilde{\beta}_0 + \sum_{i=1}^w \tilde{\beta}_{iX_i} 
\end{equation}
for every $\bm{X} \in \mathcal{I}^w$.
Let $Y$ be the indicator for whether
$\bm{X}$ is bound by the TF at chemical equilibrium. From the physical meaning of concentration,
\begin{equation}\label{eq:concentration}
{P(Y=1 \mid \bm{X})} =\frac{[F\bm{X}]}{[\bm{X}] +[F\bm{X}]}.
\end{equation}
Combining equations (\ref{eq:freeenergydef}), (\ref{eq:reftrans}) and (\ref{eq:concentration})
leads to an additive discriminant function for this free energy model (FEM), 
\begin{equation} \label{eq:logistic}
\log \frac{P(Y=1\mid \bm{X})}{P(Y=0\mid \bm{X})} = \tilde{\beta}_0 + \sum_{i=1}^w \tilde{\beta}_{iX_i}  
= \tilde{\beta}_0+\bm{X} \tilde{\bm{\beta}}:=\tilde{h}(\bm{X}).
\end{equation}
Similarly as for the WMM, we assume that $\tilde{\beta}_0$ is fixed by prior or a desired cost. 
Furthermore, it is conventional to assume that $\bm{X}$ is sampled from an
i.i.d. background model $\bm{\theta}_0$, i.e., $P(\bm{X})=\bm{\theta}_0(\bm{X})$. 
The data generation process of the FEM has a clear biological meaning. 
Suppose that we have sampled $n$ nucleotide sequences of length $w$,
$\{\bm{X}_k \in \mathcal{I}^w\}_{k=1}^n$, from the genomic background $\bm{\theta}_0$.
We mix these sequences with TF molecules in a container where the concentration
of the TF is held as a constant. At chemical equilibrium we label the sequences $\bm{X}_k$
bound by the TF as $Y_k=1$ and otherwise $Y_k=0$. The output of this experiment
is the labeled sample $\bm{D}_n=\{(Y_k,\bm{X}_k)\}_{k=1}^n$.
Although there exist other models based on binding free energy, we focus on this basic model
in this paper, which makes a theoretical analysis relatively clean while capturing main characteristics
of FE-based approaches.

Given $\bm{D}_n$, the MLE of $\tilde{\bm{\beta}}$, 
denoted by $\hat{\bm{\beta}}^{f}=\tilde{\bm{\beta}}+d\hat{\bm{\beta}}^{f}$ 
with the superscript `$f$' for FE-based estimators,
can be calculated by the standard logistic regression. 
Note that $\hat{\bm{\beta}}^{f}$ maximizes the conditional likelihood 
\begin{equation}\label{eq:FEMcondlik}
P(Y\mid \bm{X},\tilde{\bm{\beta}}) =
\frac{\exp\{(\tilde{\beta}_0+\bm{X}\tilde{\bm{\beta}})Y\} }{1+\exp(\tilde{\beta}_0+\bm{X}\tilde{\bm{\beta}}) }
\end{equation}
determined by equation (\ref{eq:logistic}).
Similar to the results in Efron (1975), 
it is not difficult to demonstrate that $\hat{\bm{\beta}}^{f}$ is  consistent for $\tilde{\bm{\beta}}$ with
asymptotic normality,
\begin{equation} \label{eq:normalbetaf}
\sqrt{n} d\hat{\bm{\beta}}^{f} \toinL \dnorm(\bm{0}, \bm{\Sigma}^f), \mbox{ as $ n\rightarrow \infty$},
\end{equation}
where $d\hat{\bm{\beta}}^{f}$ is regarded as a vector of $(J-1)w$ dimensions
(recall that $\tilde{\beta}_{is_i}=\hat{\beta}^f_{is_i}\equiv 0$ for $i=1,\cdots,w$).
The asymptotic covariance matrix
\begin{equation} \label{eq:covbetaf}
\bm{\Sigma}^f 
= \left[ \mathbb{E}_{\bm{\theta}_0} \left\{ p_1(\bm{X}) p_0(\bm{X}) \bm{C}_{\bm{X}} \bm{C}_{\bm{X}}^t \right\}  \right]^{-1}, 
\end{equation}
where $p_y(\bm{X}) = P(Y=y \mid \bm{X})$ for $y=0,1$, $\bm{C_X}$ is a $(J-1)w$-dimensional column vector coding each
$X_i$ as a factor of $J$ levels, and $\mathbb{E}_{\bm{\theta}_0}$ is taken with respect to (w.r.t.)
the background model $\bm{\theta}_0$ that generates the sequence $\bm{X}$. 

\subsection{Comparison} \label{seq:comparemodel}

Given $(\beta_0,\bm{\beta})$ in the WMM and the reference sequence $\bm{S}_{ref}$ in the FEM, 
if we let
\begin{equation}\label{eq:transfrombetas}
\tilde{\beta}_0  =  \beta_0 + \sum_{i=1}^w \beta_{is_i}, \;\;
\tilde{\beta}_{ij}   =  \beta_{ij}-\beta_{is_i}, 
\end{equation}
for $i=1,\cdots,w,\; j=1,\cdots,J$, then
the two models have the same conditional distribution $[Y\mid \bm{X}]$  
(\ref{eq:logpostoddsbybeta}, \ref{eq:logistic}) for any $\bm{X}$.
To simplify notations, we shall denote the decision function (\ref{eq:logistic}) in the FEM by 
$h(\bm{X})=\tilde{\beta}_0+\bm{X}\tilde{\bm{\beta}}$ hereafter.
Except for this condition distribution, other model assumptions are different.
The WMM assumes that the nucleotides in $\bm{X}$ are generated independently given its label $Y$. 
But this is not true for the FEM, in which the conditional
probability of $\bm{X}$ given $Y$ is
\begin{equation} \label{FEMConditional}
P(\bm{X} \mid Y, \mbox{FEM})  \propto P(Y\mid \bm{X}, \mbox{FEM}) P(\bm{X} \mid \mbox{FEM})
                  =  \frac{\exp\{(\tilde{\beta}_0+\bm{X}\tilde{\bm{\beta}})Y\} }
                  {1+\exp(\tilde{\beta}_0+\bm{X}\tilde{\bm{\beta}}) } \bm{\theta}_0(\bm{X}).
\end{equation}
Since equation (\ref{FEMConditional}) cannot be written as a product of functions of $X_i$, this model
implicitly allows dependence among $X_1,\cdots,X_w$.  
Consequently, the FEM may account for some observed nucleotide dependences
within a motif such as in Bulyk et al. (2002), Barash et al. (2003),
Zhou and Liu (2004), and Zhao et al. (2005) among others. 
On the other hand, under the FEM model the marginal distribution of $\bm{X}$ is simply the
background nucleotide distribution, i.e., $P(\bm{X} \mid \mbox{FEM})=\bm{\theta}_0(\bm{X})$, but
the marginal distribution of $\bm{X}$ under the WMM is a mixture,
\begin{equation} \label{eq:WMMMarginal}
P(\bm{X} \mid \mbox{WMM}) = q_1 \bm{\Theta}(\bm{X})+q_0 \bm{\theta}_0(\bm{X}).
\end{equation}

The different model assumptions lead to different procedures for parameter estimation, in particular
the coefficients $\bm{\beta}\,(\tilde{\bm{\beta}})$. 
As discussed in Sections~\ref{sec:WMM} and \ref{sec:FEM}, $\hat{\bm{\beta}}^m$ and $\hat{\bm{\beta}}^f$
are consistent under the WMM and under the FEM, respectively.
Since $\hat{\bm{\beta}}^f$ maximizes the conditional likelihood $P(Y\mid \bm{X},\tilde{\bm{\beta}})$ (\ref{eq:FEMcondlik})
which is identical between the two models, 
it is also consistent for $\bm{\beta}$ under the WMM up to the translation (\ref{eq:transfrombetas}).
However, $\hat{\bm{\beta}}^f$ is expected to be less efficient than $\hat{\bm{\beta}}^m$ in prediction if the WMM corresponds to
the underlying data generation process, due to
the ignorance of the information on $\bm{\Theta}$ 
contained in the marginal likelihood $P(\bm{X} \mid \bm{\Theta},\mbox{WMM})$ 
(to  be discussed in detail in Section~\ref{sec:errorWMM}). 
Conversely, if data are generated by the FEM,
$\hat{\bm{\beta}}^m$ is biased and no longer consistent. We will analyze the bias and 
the resulting incremental error rate in later sections. 

\section{Theoretical results} \label{sec:prediction}

For both WMM and FEM, the ideal decision function $h(\bm{x})$
is obtained with the true parameters of
the respective models and the corresponding ideal error rate
\begin{equation} \label{eq:idealR}
R^*=\sum_{\bm{x}: h(\bm{x})\geq 0} {P(Y=0, \bm{X}=\bm{x})}
+\sum_{\bm{x}: h(\bm{x})< 0}{P(Y=1, \bm{X}=\bm{x})}.
\end{equation}
Denote by
\begin{equation*}
R^*(\bm{x}) = \min_{y\in\{0,1\}} P(Y=y \mid \bm{X}=\bm{x})
\end{equation*}
the ideal error rate for $h(\bm{x})$ given $\bm{X}= \bm x$.
Consider a decision function $\hat{h}(\bm{x})$ estimated from $\bm{D}_n$. Given
any $\bm{x}$ for which $h(\bm{x}) \hat{h}(\bm{x}) <0$, the incremental error rate
beyond $R^*(\bm x)$ is 
\begin{equation*}
\Delta R(\bm{x}) = \left| P(Y=1\mid \bm{X}=\bm{x})- P(Y=0\mid \bm{X}=\bm{x})\right|. 
\end{equation*}
Then the expectation of the total incremental error rate for $\hat{h}$ is
\begin{eqnarray} \label{eq:EDeltaR}
\mathbb{E}[\Delta R(\hat{h})]  &= &
 \mathbb{E} \left[ \sum_{\bm{x} \in \mathcal{I}^w} \Delta R(\bm{x}) P(\bm{X}=\bm{x})\, 
 \bm{1}\left\{ h(\bm{x}) \hat{h}(\bm{x}) <0 \right\} \right] \notag \\ 
 & = & \sum_{\bm{x} \in \mathcal{I}^w} \Delta R(\bm{x}) P(\bm{X}=\bm{x}) P\left\{ h(\bm{x}) \hat{h}(\bm{x}) <0 \right\},
\end{eqnarray}
where $\bm{1}(\cdot)$ is the indicator function.
Please note that $\hat{h}$, constructed from a sample of size $n$, is a random function. 
Let $\Delta \hat{h}(\bm{x}) =  \hat{h}(\bm{x})-h(\bm{x})$ be the deviation of  $\hat{h}(\bm{x})$ from $h(\bm{x})$.
In what follows, we will derive two theorems on $\mathbb{E}[\Delta R(\hat{h})]$ under
different assumptions for $\Delta \hat{h}(\bm{x})$. 
As we will see, the asymptotic error rates of the WM and the FE procedures
under the data generation models discussed in this paper can all be calculated based on the two theorems.

Suppose that, for every $\bm{x}$, $\sqrt{n} \Delta \hat{h}(\bm{x}) \toinL \dnorm(0, V(\hat{h},\bm{x}))$, where
$V(\hat{h},\bm{x})$ is the asymptotic variance. 
As $n\to \infty$,
\begin{eqnarray} \label{eq:EDeltaRasym}
\mathbb{E}[ \Delta R(\hat{h})]  
& \to &  \sum_{\bm{x} \in \mathcal{I}^w}  \Delta R(\bm{x}) P(\bm{X}=\bm{x}) 
			P(\Delta \hat{h}(\bm{x}) > |h(\bm{x})|) \notag \\
         & = & \sum_{\bm{x} \in \mathcal{I}^w}  \Delta R(\bm{x}) P(\bm{X}=\bm{x}) \;
                                \Phi \left\{-  \sqrt{n h^2(\bm{x}) /V(\hat{h},\bm{x})} \right\},
\end{eqnarray}
where $\Phi$ is the cdf of the standard normal distribution $\dnorm(0,1)$.
Let
\begin{equation} \label{eq:alphah}
\alpha(\hat{h})=\min_{h(\bm{x})\neq 0}h^2(\bm{x})/V(\hat{h},\bm{x}),
\end{equation}
and $\bm{x}^*$ be the corresponding minimum.
Note that $\Delta R(\bm{x})=0$ when $h(\bm{x})=0$.
Thus, as $n\rightarrow \infty$, $\mathbb{E}[ \Delta R(\hat{h})]  $ is dominated by the term 
$ \Delta R(\bm{x}^*) P(\bm{X}=\bm{x}^*)\; \Phi \left[- \{n \alpha(\hat{h})\}^{1/2} \right]$,
where $\alpha(\hat{h})$ determines the rate of convergence. 
Using the theory of large deviations, we obtain: 
\begin{theorem} \label{th:ConsistentIncRate}
If $\sqrt{n} \Delta \hat{h}(\bm{x})  \toinL \dnorm(0, V(\hat{h},\bm{x}))$ for every $\bm{x} \in \mathcal{I}^w$ then 
\begin{equation*}
\frac{1}{n}\log \mathbb{E}[ \Delta R(\hat{h})]  \rightarrow -\frac{\alpha(\hat{h})}{2}, \mbox{ as $n\rightarrow \infty$}.
\end{equation*}
\end{theorem}

Let $\hat{h}^a$ and $\hat{h}^b$ be two estimated decisions 
constructed from samples of size $n_a$ and $n_b$, respectively. Suppose that both of them satisfy the condition
in Theorem~\ref{th:ConsistentIncRate}. We define the {\em asymptotic relative efficiency} (ARE) 
of $\hat{h}^a$ with respect to $\hat{h}^b$
 by $\mbox{ARE}(\hat{h}^a,\hat{h}^b)=\alpha(\hat{h}^a)/\alpha(\hat{h}^b)$,
 which is the limit ratio $n_b/n_a$ required to achieve the same asymptotic performance.

If $\hat{h}(\bm{x})$ is biased in the sense that
$\sqrt{n} \{\Delta \hat{h}(\bm{x}) -\mu(\hat{h},\bm{x}) \} \toinL \dnorm(0, V(\hat{h},\bm{x}))$,
where $\mu(\hat{h},\bm{x})$ denotes the asymptotic bias of $\hat{h}(\bm{x})$,
then simple derivation from equation (\ref{eq:EDeltaR}) gives that
\begin{equation*} 
\mathbb{E}[ \Delta R(\hat{h})]  
 \to  \sum_{\bm{x} \in \mathcal{I}^w}  \Delta R(\bm{x}) P(\bm{X}=\bm{x}) \;
     \Phi \left[ \frac{-\sqrt{n} \sign\{h(\bm{x})\} \{h(\bm{x}) + \mu(\hat{h}, \bm{x} )\} } {\sqrt{V(\hat{h},\bm{x})} } \right],
\end{equation*}
as $n\to \infty$, where $\mbox{sign}(y)$ is the sign of $y$ with $\mbox{sign}(0)\equiv 0$.
\begin{theorem} \label{th:BiasedIncRate}
Suppose that $\sqrt{n} \{\Delta \hat{h}(\bm{x}) -\mu(\hat{h},\bm{x}) \} \toinL \dnorm(0, V(\hat{h},\bm{x}))$ for every $\bm{x}\in \mathcal{I}^w$. 
Let $\mathcal{B}(\hat{h}) = \{ \bm{x} : \emph{sign}\{h(\bm{x})\} \{h(\bm{x}) + \mu(\hat{h}, \bm{x} )\} < 0 \}$. Then
\begin{equation}\label{eq:EDeltaRasymbias}
\mathbb{E}[ \Delta R(\hat{h})]  \rightarrow  \sum_{\bm{x} \in \mathcal{B}(\hat{h}) }  
 \Delta R(\bm{x}) P(\bm{X}=\bm{x}), \mbox{ as $n\rightarrow \infty$.}
\end{equation}
\end{theorem}
We ignore the case $\{\bm{x}: h(\bm{x}) + \mu(\hat{h}, \bm{x} ) =0\}$ which practically never happens.
The set $\mathcal{B}(\hat{h})$ is the collection of $\bm{x}$ for which the estimated decision $\hat{h}$ gives a
different predicted label from the ideal decision $h$ as $n\to \infty$.
Note that $\mathbb{E}[ \Delta R(\hat{h})] $ does not vanish if $\mathcal{B}(\hat{h})$ is nonempty.
Thus, the incremental percentage over the ideal error rate, $\mathbb{E}[ \Delta R(\hat{h})] / R^*$,
is an appropriate measure of the predictive performance of $\hat{h}$.

In the remainder of this section, we derive and compare the error rates of the WM and the FE procedures.
From Sections~\ref{sec:errorWMM} to \ref{sec:FEMMCbg}, we assume that
the constant term $\beta_0 (\tilde{\beta}_0)$ is fixed to its true value. The results are generalized
to situations where the constant is mis-specified in Section~\ref{sec:constant}.
The computation of $\alpha(\hat{h})$ (\ref{eq:alphah}) and $\mathbb{E}[ \Delta R(\hat{h})]$ (\ref{eq:EDeltaRasymbias})
will be discussed in Section~\ref{sec:computation}.

\subsection{Error rates under WMM} \label{sec:errorWMM}

In this subsection we assume that the underlying data generation process is given by the WMM.
Since both $\hat{\bm{\beta}}^m$ and $\hat{\bm{\beta}}^f$ are consistent with 
asymptotic normality under the WMM,
we may uniformly denote their decision functions 
by $\hat{h}(\bm{x}) =\beta_0 + \bm{x}\hat{\bm{\beta}}= h(\bm{x}) + \bm{x} d \hat{\bm{\beta}}$,
where $d \hat{\bm{\beta}}=\hat{\bm{\beta}}-\bm{\beta}$ and 
$\sqrt{n} d\hat{\bm{\beta}}$ follows a normal distribution $\dnorm(\bm{0},\bm{\Sigma})$ as $n\to \infty$.
This implies that 
$\sqrt{n} \Delta \hat{h}(\bm{x})=\sqrt{n} \bm{x} d \hat{\bm{\beta}} \toinL \dnorm(0, V^m(\hat{\bm{\beta}},\bm{x}))$
with $V^m(\hat{\bm{\beta}},\bm{x})$ being the asymptotic variance. 
The superscript `$m$' indicates the WMM as the data generation model.
Let $\mathbb{E}[ \Delta R^m(\hat{\bm{\beta}})] $ be the  expected incremental error rate
of $\hat{h}$ indexed by $\hat{\bm{\beta}}$.
Following Theorem~\ref{th:ConsistentIncRate},
\begin{equation} \label{eq:alpham}
\frac{1}{n} \log \mathbb{E}[\Delta R^m(\hat{\bm{\beta}})] \to -\frac{\alpha^m(\hat{\bm{\beta}})}{2}
= -\frac{1}{2}\min_{h(\bm{x})\neq 0} \frac{h^2(\bm{x})}{V^m(\hat{\bm{\beta}},\bm{x})},
\end{equation}
as $n\rightarrow \infty$. 
Consequently, the ARE of the FE procedure w.r.t  the WM procedure, 
$\mbox{ARE}^m(\hat{\bm{\beta}}^f,\hat{\bm{\beta}}^m)$, is determined
by the ratio of $\alpha^m(\hat{\bm{\beta}}^f)$ over $\alpha^m(\hat{\bm{\beta}}^m)$.


The decision function of the WM procedure is constructed with $\hat{\bm{\beta}}^m$ (Section~\ref{sec:WMM}).
Note that $\bm{x} d \hat{\bm{\beta}}=\sum_i d\hat{\beta}_{ix_i}$ is a summation of $w$ $d\hat{\beta}_{ij}$'s,
each from a different $d\hat{\bm{\beta}}_i$. The limiting distribution of $\sqrt{n}d\hat{\beta}^m_{ij}$ 
(\ref{eq:limitingbetam}) and the mutual independence among $d\hat{\bm{\beta}}^m_{i}$ 
imply that the asymptotic variance of $\sqrt{n} \bm{x}d\hat{\bm{\beta}}^m$ is
\begin{equation*}
\frac{1}{q_1} \sum_{i=1}^w (1-\theta_{ix_i}) / \theta_{ix_i}= \frac{1}{q_1}\bm{x} \left\{ (1-\bm{\Theta})/\bm{\Theta}  \right\}, 
\end{equation*}
and consequently, 
\begin{equation*}
\alpha^m(\hat{\bm{\beta}}^m)  =\min_{\bm{x\beta}\neq -\beta_0} 
\frac{q_1 (\beta_0+\bm{x\beta})^2}{ \bm{x} \{(1-\bm{\Theta})/\bm{\Theta}\}}.
\end{equation*}


Suppose that we have chosen $(s_1,\cdots,s_w)$ as the reference sequence in
the FE procedure. 
Define $\tilde{{\beta}_0}$ and 
$\tilde{\bm{\beta}}$ from the parameters $(\beta_0,\bm{\beta})$ of the WMM by equation (\ref{eq:transfrombetas}).
Then the FE-based estimator $\hat{\bm{\beta}}^f$ is
consistent for $\tilde{\bm{\beta}}$ with asymptotic normality. 
Let $d\hat{\bm{\beta}}^f=\hat{\bm{\beta}}^f-\tilde{\bm{\beta}}$. Similar to equation (\ref{eq:covbetaf}),
the asymptotic covariance matrix of $\sqrt{n}d\hat{\bm{\beta}}^f$ is
$
\left[ \mathbb{E} \left\{ p_1(\bm{X}) p_0(\bm{X}) \bm{C}_{\bm{X}} \bm{C}_{\bm{X}}^t \right\} \right]^{-1} ,
$
where the expectation is taken w.r.t. the marginal distribution of $\bm{X}$ under the WMM (\ref{eq:WMMMarginal}).
Thus the covariance matrix can be written as
\begin{equation} \label{eq:covbetafwmm}
\mbox{Cov}^m\left( \sqrt{n} d\hat{\bm{\beta}}^{f} \right)
=  \left[ q_0 \mathbb{E}_{\bm{\theta}_0} \left\{ \frac{ e^{h(\bm{X})} }{e^{h(\bm{X})}+1} \bm{C}_{\bm{X}} \bm{C}_{\bm{X}}^t \right\} \right]^{-1} ,
\end{equation}
where the expectation $\mathbb{E}_{\bm{\theta}_0}$ averages over $\bm{X}\in \mathcal{I}^w$ generated from 
the background model $\bm{\theta}_0$.
Based on equation (\ref{eq:covbetafwmm}), one can calculate the variance of 
$\sqrt{n} \bm{x}d\hat{\bm{\beta}}^f$ for every $\bm{x}$ and determine the convergence rate 
$\alpha^m(\hat{\bm{\beta}}^f)$ of the expected incremental error rate $\mathbb{E}[ \Delta R^m(\hat{\bm{\beta}}^f)]$
for the FE procedure.

Because the estimation of $\hat{\bm{\beta}}^f$ is only based on the conditional distribution $[Y\mid \bm{X}]$
while $\hat{\bm{\beta}}^m$ is estimated from the joint distribution of $Y$ and $\bm{X}$,
we expect $\hat{\bm{\beta}}^f$ to be less efficient in prediction with 
$\alpha^m(\hat{\bm{\beta}}^f)<\alpha^m(\hat{\bm{\beta}}^m)$.
We will conduct a numerical study in Section~\ref{sec:numericalWMM} to 
evaluate $\mbox{ARE}^m(\hat{\bm{\beta}}^f,\hat{\bm{\beta}}^m)$
on 200 transcription factors to confirm our conclusion.
Here we demonstrate the lower efficiency of $\hat{\bm{\beta}}^f$ by the loss of
Fisher information in estimating an individual $\theta_{ij}$ from the conditional likelihood only.
For simplicity, suppose that $\bm{\Theta}_{[-i]}$ is given and collapse $X_i$ into two categories, 
$X_i=j$ and $X_i\neq j$. Because 
\begin{equation*}
P(\bm{X},Y \mid \bm{\Theta} )= P(Y \mid \bm{X}, \bm{\Theta}) P(\bm{X} \mid \bm{\Theta}) 
\end{equation*}
under the WMM, 
the loss of information equals the Fisher information on $\theta_{ij}$ contained in 
the marginal likelihood $P(\bm{X} \mid \bm{\Theta})$,
denoted by $I(\theta_{ij} \mid \bm{X})$.
Let $I(\theta_{ij} \mid \bm{X},Y)$ be the Fisher information on $\theta_{ij}$ given $\bm{X}$ and $Y$ jointly.
We define 
\begin{equation} \label{eq:deflossinfo}
\Delta (\theta_{ij} \mid [Y\mid \bm{X}]) = I(\theta_{ij} \mid \bm{X}) / I(\theta_{ij} \mid \bm{X},Y)
\end{equation}
as the fraction of the loss of information on $\theta_{ij}$ in the conditional likelihood $P(Y \mid \bm{X}, \bm{\Theta})$.
\begin{proposition} \label{lm:lossinfo}
Let $\bar{\theta}_{ij} = q_0 \theta_{0j} + q_1 \theta_{ij}$.
If $(Y,\bm{X})$ is drawn from the WMM then 
\begin{equation*} 
\Delta (\theta_{ij} \mid [Y\mid \bm{X}]) \geq \frac{q_1\theta_{ij}(1-\theta_{ij})}{\bar{\theta}_{ij}(1-\bar{\theta}_{ij})}
:= B (q_1,{\theta}_{ij},{\theta}_{0j}).
\end{equation*}
\end{proposition}
A proof of this proposition is given in Appendix A.
If one chooses to include an equal number of background sites ($Y=0$) and binding sites ($Y=1$)
in logistic regression to estimate $\hat{\bm{\beta}}^f$, which effectively specifies
$q_0 = q_1=0.5$ by design, then this lower bound may be substantial.
For example, with a uniform background distribution 
$\theta_{0j} = 0.25$ for $j=1,\cdots,4$, 
the range of $B (q_1,{\theta}_{ij},{\theta}_{0j})  $ is between 20\% and 55\% for
most typical values of $\theta_{ij}$ (Table~\ref{tab:lowerbound}).

\begin{table}[ht]
\begin{center}
  \caption{Typical values of $B(0.5,\theta_{ij},0.25)$ \label{tab:lowerbound}}
  \vspace{0.2cm}
   \begin{tabular}{ccccccccccc} 
   \hline
   $\theta_{ij}$ & 0.1 & 0.2 & 0.3 &0.4 & 0.5 & 0.6 & 0.7 & 0.8 & 0.9 \\
   $B(0.5,\theta_{ij},0.25)$(\%) & 31 & 46 & 53 & 55 & 53 & 49 & 42 & 32 & 18 \\
    \hline
  \end{tabular}
   \end{center}
   \end{table}

\subsection{WMM with Markov background}\label{sec:WMMwithMCbg}

We generalize the background model to a Markov chain,
which often represents a better fit to genomic background in high organisms.
We assume that given $Y=0$, $\bm{X}$ is generated by a first order Markov chain
with a transition probability matrix $\bm{\psi}_0=(\psi_0(x,y))_{J\times J}$ where $x,y \in \mathcal{I}$.
For any $\bm{x}=(x_1,\cdots,x_w) \in \mathcal{I}^w$, $\bm{\psi}_0(\bm{x}) := \prod_{i=1}^w \psi_0(x_{i-1},x_i)$,
where $\psi_0(x_0,x_1)$ is interpreted as the probability of $x_1$ under the stationary distribution
of the Markov chain. The ideal decision function under this model is 
\begin{equation}\label{eq:decisionWMMMC}
h_1(\bm{x})=\log\frac{P(Y=1\mid \bm{X}=\bm{x})}{P(Y=0\mid \bm{X}=\bm{x})} = \beta_0 + 
\sum_{i=1}^w \log\theta_{ix_i} -\log \psi_0(x_{i-1},x_i),
\end{equation}
where $\beta_0=\log(q_1/q_0)$ and the subscript `1' [in  $h_1(\bm x)$ and $\Delta R_1^m$ (\ref{eq:EDeltaRmfMC})]
indicates a quantity whose definition involves a Markov background model.
Since $\bm{\psi}_0$ can be accurately estimated with sufficient genomic background sequences,
we assume that it is given. With the MLE $\hat{\bm{\Theta}}^m$ from observed binding sites, 
the WM procedure constructs 
a decision whose expected incremental error rate converges to zero exponentially fast 
as $n\rightarrow \infty$, following Theorem~\ref{th:ConsistentIncRate}.

With a slight abuse of notations,
let us denote by $\bm{\theta}_0$ the probability vector of the stationary distribution of the Markov chain,
which is also the marginal distribution of any nucleotide $X_i$ in a background site. 
We still define $\beta_0$ and $\bm{\beta}$ by equation (\ref{eq:definebeta}) with $\bm{\theta}_0$
being the stationary probabilities, and translate $\beta_0$ and $\bm{\beta}$ via
a reference sequence to $\tilde{\beta}_0$ and $\tilde{\bm{\beta}}$ (\ref{eq:transfrombetas}).
Let $(Y,\bm{X})$ be a sample from the WMM with Markov background.
If the dependence among neighboring nucleotides in a background site is ignored,
the conditional likelihood $P(Y\mid \bm{X}, \tilde{\bm{\beta}})$, parameterized by $\tilde{\bm{\beta}}$,
is then given by the same expression in equation (\ref{eq:FEMcondlik}).
Because the FE-based estimator $\hat{\bm{\beta}}^f$ maximizes
this conditional likelihood, it is standard to show that $\hat{\bm{\beta}}^f \toinP \tilde{\bm{\beta}}$ and
is asymptotically normal. Let $\hat{h}^f(\bm{x})=\tilde{\beta}_0+ \bm{x} \hat{\bm{\beta}}^f$ denote
the estimated decision function of the FE procedure.
As $n\to \infty$,
\begin{equation} \label{eq:hfWMMMC}
\hat{h}^f(\bm{x}) \toinP \tilde{\beta}_0+ \bm{x} \tilde{\bm{\beta}}=\beta_0+\sum_{i=1}^w \log\theta_{ix_i}-\log\theta_{0x_i}\,.
\end{equation}
Let $\Delta \hat{h}^f(\bm{x}) = \hat{h}^f(\bm{x})-h_1(\bm{x})$ be the deviation of $\hat{h}^f(\bm{x})$
from the ideal decision (\ref{eq:decisionWMMMC}). 
Comparing equations (\ref{eq:decisionWMMMC}) and (\ref{eq:hfWMMMC}) gives the asymptotic bias,
\begin{equation*}
b(\bm{x})=\sum_{i=1}^w \log \psi_0(x_{i-1},x_i)-\log\theta_{0x_i} = \log\bm{\psi}_0(\bm{x}) -\log\bm{\theta}_0(\bm{x}).
\end{equation*}
Due to the asymptotic normality of $\hat{\bm{\beta}}^f$,
we have $\sqrt{n} \{\Delta \hat{h}^f(\bm{x})- b(\bm{x})\} \toinL \dnorm(0, V_1^m(\hat{\bm{\beta}}^f,\bm{x}))$,
where $V_1^m(\hat{\bm{\beta}}^f,\bm{x})$ is the corresponding asymptotic variance.
Under this model, 
\begin{equation*}
\Delta R(\bm{x}) P(\bm{X}=\bm{x})  =   \left|  q_1 \bm{\Theta}(\bm{x})  - q_0 \bm{\psi}_0 (\bm{x}) \right| 
 =  q_0 \bm{\psi}_0 (\bm{x})\, \left| e^{h_1(\bm{x})}-1 \right|. 
\end{equation*}
Following Theorem~\ref{th:BiasedIncRate} with $\mu(\hat{h}^f,\bm{x})=b(\bm{x})$, 
the expected incremental error rate
\begin{equation}\label{eq:EDeltaRmfMC}
 \mathbb{E}[\Delta R_1^m(\hat{\bm{\beta}}^f)]   \rightarrow  \sum_{ \mathcal{B}^m_1(\hat{\bm{\beta}}^f)}  
 q_0\, \left| e^{h_1(\bm{x})}-1 \right| \; \bm{\psi}_0 (\bm{x}), \mbox{ as $n\rightarrow \infty$},
\end{equation}
where $\mathcal{B}^m_1(\hat{\bm{\beta}}^f) = \{ \bm{x} : \mbox{sign}\{h_1(\bm{x})\} \{h_1(\bm{x}) + b(\bm{x})\} < 0 \}$. 
The incremental percentage over the ideal error rate, $\mathbb{E}[\Delta R_1^m(\hat{\bm{\beta}}^f)] /(R^{m}_1)^*$,
is appropriate for comparing the FE-based prediction with the WM-based prediction whose
expected error rate converges to $(R^{m}_1)^*$. 
A general expression for $R^*$ is given in equation (\ref{eq:idealR}) which, under the WMM with Markov background, 
is written as
\begin{equation} \label{eq:IdErrWMMMC} 
(R^{m}_1)^*=q_0 \mathbb{E}_{\bm{\psi}_0} \left\{ {\bm{1}(h_1(\bm{X}) \geq 0)} +e^{h_1(\bm{X})} {\bm{1}(h_1(\bm{X}) < 0)}\right\}.
\end{equation}

\subsection{Error rates under FEM} \label{sec:errorFEM}

We now analyze asymptotic error rates of the two procedures regarding
the FEM as the underlying data generation mechanism.

The FE-based estimator $\hat{\bm{\beta}}^f$ is consistent for $\tilde{\bm{\beta}}$ under the FEM.
The asymptotic normality of $\sqrt{n}d \hat{\bm{\beta}}^f$ (\ref{eq:normalbetaf}, \ref{eq:covbetaf}) 
implies that $\sqrt{n}\bm{x} d \hat{\bm{\beta}}^f \toinL \dnorm(0, V^f(\hat{\bm{\beta}}^f,\bm{x}))$.
Let $\mathbb{E}[ \Delta R^f(\hat{\bm{\beta}}^f)]$ be the expected incremental error rate of the
FE procedure under the FEM. 
From Theorem~\ref{th:ConsistentIncRate}, we have
\begin{equation} \label{eq:FEFEMerror}
\frac{1}{n}\log \mathbb{E}[ \Delta R^f(\hat{\bm{\beta}}^f)]  \rightarrow 
-\frac{\alpha^f(\hat{\bm{\beta}}^f)}{2}=-\frac{1}{2} \min_{h(\bm{x})\neq 0} \frac{h^2(\bm{x})}{V^f(\hat{\bm{\beta}}^f,\bm{x})}, 
\mbox{ as $n\rightarrow \infty$}.
\end{equation}

Denote by $\bm{\theta}^f_{i}=(\theta^f_{i1},\cdots,\theta^f_{iJ})$ the probability vector of 
the conditional distribution $[X_i \mid Y=1]$ under the FEM, i.e.,
\begin{equation} \label{eq:thetaFEM}
\theta^f_{ij}=P(X_i=j\mid Y=1, \mbox{FEM}),
\end{equation}
for $i=1,\cdots,w$, and call $\bm{\Theta}^f=(\theta^f_{ij})_{w \times J}$ the weight matrix.
Recall that the WM-based estimator $\hat{\bm{\beta}}^m$ is obtained by estimating $\bm{\theta}^f_i$
individually from observed binding sites $\bm{D}_n^+$ and then transforming the estimates via equation
(\ref{eq:definebeta}). Denote the estimated weight matrix by $\hat{\bm{\Theta}}^m$. 
Since the data are generated by the FEM, $\hat{\bm{\Theta}}^m \toinP \bm{\Theta}^f$ and
$\sqrt{n}d\hat{\bm{\Theta}}^m=\sqrt{n}(\hat{\bm{\Theta}}^m-\bm{\Theta}^f)$ follows a multivariate normal distribution
asymptotically, similar to (\ref{eq:dthetamdistrn}), 
but $d\hat{\bm{\theta}}_i^m$  and $d\hat{\bm{\theta}}_k^m$ may be correlated ($1\leq i \neq k \leq w$).
Given that the coefficients $\tilde{\bm{\beta}}$ in the FEM are defined w.r.t. a reference sequence,
we transform $\hat{\bm{\Theta}}^m$ to 
\begin{equation} \label{eq:betamFEM}
\hat{\beta}^m_{ij} = \log(\hat{\theta}^m_{ij}/\theta_{0j}) -\log (\hat{\theta}^m_{is_i}/\theta_{0s_i}), \mbox{ for all $i,j$},
\end{equation}
where $s_i$ is the $i$th nucleotide of the reference sequence $\bm{S}_{ref}$. 
Let $\Delta \hat{\bm{\beta}}^m=\hat{\bm{\beta}}^m-\tilde{\bm{\beta}}$ be the deviation of 
$\hat{\bm{\beta}}^m=(\hat{\beta}^m_{ij})_{w\times J}$.
To obtain its asymptotic distribution, we determine the cell probability $\theta^f_{ij}$ (\ref{eq:thetaFEM})
from equation (\ref{FEMConditional}), that is,
\begin{equation*}
\theta^f_{ij}  \propto  \theta_{0j} e^{\tilde{\beta}_{ij}} \sum_{\bm{x} \in \mathcal{I}^{w-1}} 
			\frac{e^{\tilde{\beta}_0+\bm{x}\tilde{\bm{\beta}}_{[-i]}} }{1+e^{\tilde{\beta}_{ij}}e^{\tilde{\beta}_0+\bm{x}\tilde{\bm{\beta}}_{[-i]}} } 
			\bm{\theta}_0(\bm{x}) 
		 =  \theta_{0j} e^{\tilde{\beta}_{ij}} \mathbb{E}_{\bm{\theta}_0}\left\{ \frac{e^{U_i}}{1+e^{\tilde{\beta}_{ij}} e^{U_i} } \right\},
\end{equation*}
where $U_i=\tilde{\beta}_0+\bm{X}\tilde{\bm{\beta}}_{[-i]}$ for $\bm{X} \in \mathcal{I}^{w-1}$.
In particular, $\theta^f_{is_i} \propto \theta_{0s_i}  \mathbb{E}_{\bm{\theta}_0}\left\{ {e^{U_i}}/{(1+e^{U_i}) } \right\}$
since $\tilde{\beta}_{is_i}=0$.
For $i=1,\cdots,w$ and $j=1,\cdots,J$, we define
\begin{equation} \label{eq:deltaij}
\delta_{ij}=\log \mathbb{E}_{\bm{\theta}_0}\left\{ \frac{e^{U_i}}{1+e^{\tilde{\beta}_{ij}} e^{U_i} } \right\}
-\log \mathbb{E}_{\bm{\theta}_0}\left\{ \frac{e^{U_i}}{1+e^{U_i} } \right\}
\end{equation}
and rewrite $\theta^f_{ij} =  \theta_{0j} e^{\tilde{\beta}_{ij}+\delta_{ij}}/Z_i$,
where $Z_i=\sum_j \theta_{0j} e^{\tilde{\beta}_{ij}+\delta_{ij}}$ is the normalizing constant.
Because $\hat{\theta}^m_{ij}\toinP {\theta}^f_{ij}$ and $\tilde{\beta}_{is_i}=\delta_{is_i}=0$, 
from equation (\ref{eq:betamFEM}) we have
\begin{equation} \label{eq:betamFEMinP}
\hat{\beta}^m_{ij}\toinP \log(\theta^f_{ij}/\theta_{0j})-\log(\theta^f_{is_i}/\theta_{0s_i})=\tilde{\beta}_{ij}+\delta_{ij}
\end{equation}
for all $i$ and $j$ as $n\rightarrow \infty$. 
Thus, $\bm{\delta} = (\delta_{ij})_{w\times J}$ is the asymptotic bias of $\hat{\bm{\beta}}^m$. 
From the asymptotic normality of $\sqrt{n}d\hat{\bm{\Theta}}^m$, we see that
$\sqrt{n} (\Delta \hat{\bm{\beta}}^m-\bm{\delta})$ follows a multivariate normal distribution
with mean $\bm{0}$ and finite covariance matrix as $n\rightarrow \infty$.
Note that this multivariate normal distribution is defined on a $(J-1)w$-dimensional space,
since $\Delta \hat{{\beta}}_{is_i}^m=\delta_{is_i} \equiv 0$ for $i=1,\cdots,w$.

Consider the WM-based decision function
$\hat{h}^m(\bm{x}) = \tilde{\beta}_0+\bm{x} \hat{\bm{\beta}}^m= h(\bm{x}) + \bm{x}\Delta \hat{\bm{\beta}}^m$.
The above derivation shows that 
$\sqrt{n} (\bm{x}\Delta \hat{\bm{\beta}}^m-\bm{x} \bm{\delta}) \toinL \dnorm(0, V^f(\hat{\bm{\beta}}^m,\bm{x}))$,
where the variance is determined by the covariance matrix of $\Delta \hat{\bm{\beta}}^m$.
Following Theorem~\ref{th:BiasedIncRate}, the expectation of the total incremental error rate of the WM procedure 
\begin{equation} \label{eq:EDeltaRfm}
 \mathbb{E}[ \Delta R^f(\hat{\bm{\beta}}^m)]   \rightarrow  
 \sum_{\mathcal{B}^f(\hat{\bm{\beta}}^m)} \tanh | h(\bm{x})/2 | \; \bm{\theta}_0 (\bm{x}),
 \mbox{ as $n\rightarrow \infty$},
 \end{equation}
where $\mathcal{B}^f(\hat{\bm{\beta}}^m) = \{ \bm{x} : \mbox{sign}\{h(\bm x)\} \{h(\bm x) + \bm{x} \bm{\delta}\} < 0 \}$.
Similarly, the incremental percentage over the ideal error rate $\mathbb{E}[ \Delta R^f(\hat{\bm{\beta}}^m)]/(R^f)^*$
is used to compare the predictions of the WM and the FE procedures, given that
$\mathbb{E}[\Delta R^f(\hat{\bm{\beta}}^f)] \to 0$ (\ref{eq:FEFEMerror}). 
Under the FEM, the ideal error rate
\begin{equation} \label{eq:FEidealrate}
(R^f)^*=\mathbb{E}_{\bm{\theta}_0} \left\{ p_0(\bm{X}){\bm{1}(h(\bm{X}) \geq 0)} +p_1(\bm{X}){\bm{1}(h(\bm{X}) < 0)}\right\}.
\end{equation}
Recall that $p_y(\bm{X})=P(Y=y \mid \bm{X})$, i.e.,
\begin{equation*}
p_y(\bm{X})= \frac{e^{yh(\bm{X})}}{e^{h(\bm{X})}+1}, \mbox{ for $y=0,1$}.
\end{equation*}

\subsection{FEM with Markov background}\label{sec:FEMMCbg}

Next, we generalize the FEM to Markov background and
assume that any sequence $\bm{X} \in \mathcal{I}^w$ is generated marginally by a Markov chain.
Consistent with Section~\ref{sec:WMMwithMCbg}, we denote by $\bm{\psi}_0=(\psi_0(x,y))_{J\times J}$ 
the transition probability matrix of the Markov chain. 

It is trivial to see that, with the Markov background model, the ideal decision is still 
$h(\bm{x})=\tilde{\beta}_0+\bm{x}\tilde{\bm{\beta}}$.
If $V_1^f(\hat{\bm{\beta}}^f,\bm{x})$ denotes the asymptotic variance of $\sqrt{n}\bm{x} d\hat{\bm{\beta}}^f$
under Markov background, then with $V_1^f$ in place of $V^f$ 
equation (\ref{eq:FEFEMerror}) remains valid for the FE-based prediction.
On the other hand, if we proceed with the WM procedure, 
the expected incremental error rate
\begin{equation}\label{eq:WMErrinFEMMC} 
\mathbb{E}[ \Delta R_1^f(\hat{\bm{\beta}}^m)]\to  
\sum_{ \mathcal{B}_1^f(\hat{\bm{\beta}}^m)} \tanh | h(\bm{x})/2 | \; \bm{\psi}_0(\bm{x}),
 \mbox{ as $n\rightarrow \infty$},
\end{equation}
where $\mathcal{B}_1^f(\hat{\bm{\beta}}^m) = \{ \bm{x} : \mbox{sign}\{h(\bm x)\} \{h(\bm x) + \delta(\bm{x})\} < 0 \}$.
Here $\delta(\bm{x})$ denotes the asymptotic bias of the WM-based decision function for $\bm{x}$. A detailed
derivation of equation (\ref{eq:WMErrinFEMMC}) and the bias $\delta(\bm{x})$ is provided in Appendix B.
In analogy to the FEM with i.i.d. background, $\mathbb{E}[ \Delta R_1^f(\hat{\bm{\beta}}^m)]/(R_1^f)^*$ measures
the increased error rate of the WM procedure relative to the FE procedure,
where 
\begin{equation*}
(R_1^f)^*=\mathbb{E}_{\bm{\psi}_0} \left\{ p_0(\bm{X}){\bm{1}(h(\bm{X}) \geq 0)} +p_1(\bm{X}){\bm{1}(h(\bm{X}) < 0)}\right\}.
\end{equation*}

\subsection{Mis-specification of the constant term}\label{sec:constant}

In all the above derivations, we have assumed that the constant term $\beta_0 (\tilde{\beta}_0)$ is fixed to its true value. If
this is not the case, then the deviation $\Delta \hat{\beta}_0= \hat{\beta}_0 -\beta_0 (\tilde{\beta}_0)$
will be an extra bias term for an estimated decision in which the constant term is fixed to $\hat{\beta}_0$.
More specifically, the set 
$\mathcal{B}^f(\hat{\bm{\beta}}^m)$ in equation (\ref{eq:EDeltaRfm}) will be replaced by 
$\{ \bm{x} : \mbox{sign}\{h(\bm x)\} \{h(\bm x) + \bm{x} \bm{\delta} + \Delta \hat{\beta}_0\} < 0 \}$, and
similarly for $\mathcal{B}_1^m(\hat{\bm{\beta}}^f)$ in equation (\ref{eq:EDeltaRmfMC}) and 
$\mathcal{B}_1^f(\hat{\bm{\beta}}^m)$ in equation (\ref{eq:WMErrinFEMMC}).

\section{Computation} \label{sec:computation}

To apply the theoretical results, we need to solve the minimization (\ref{eq:alphah}) 
and the summation (\ref{eq:EDeltaRasymbias})
involved in Theorems~\ref{th:ConsistentIncRate} and \ref{th:BiasedIncRate}, respectively.
If the width of a motif $w\leq 12$, brute-force enumeration of all $w$-mers 
is computationally feasible, which provides exact solutions for both the minimization and
the summation problems.

For a motif of width $w>12$, we minimize (\ref{eq:alphah}) to find $\alpha(\hat{h})$
by a two-step approach. We
generate $N=5\times 10^6$ $w$-mers from the background model $\bm{\theta}_0$
and identify the minimum of  (\ref{eq:alphah}) among them.
Then we refine the obtained minimum by simulated annealing for 5,000 iterations
with temperature decreasing linearly from one to zero. 
At each iteration, we randomly choose one nucleotide $X_i$ from the $w$ positions
and propose to mutate $X_i$ to one of the other three nucleotide bases with equal probability.
The proposal is accepted according to a Metropolis-Hastings ratio with current temperature.

Since the set $\mathcal{B}(\hat{h})$, as in equations (\ref{eq:EDeltaRmfMC}), (\ref{eq:EDeltaRfm}) and (\ref{eq:WMErrinFEMMC}), 
is usually small, it will be very inefficient to approximate the summation by generating $w$-mers from background distributions. Thus,
we develop an importance sampling approach to approximate the summation (\ref{eq:EDeltaRasymbias}) when $w>12$.
Here, we use the calculation of $\mathbb{E}[ \Delta R^f(\hat{\bm{\beta}}^m)]$ (\ref{eq:EDeltaRfm}) to illustrate this approach.
Note that one can bound $\bm{x\delta}$ in the definition of $\mathcal{B}^f(\hat{\bm{\beta}}^m)$ so that
$\bm{x\delta} \in [M_1,M_2]$, where $M_1=\sum_{i=1}^w \min_j \delta_{ij}$ and $M_2=\sum_{i=1}^w \max_j \delta_{ij}$.
These bounds imply that if $\bm{x} \in \mathcal{B}^f(\hat{\bm{\beta}}^m)$ then
\begin{equation*}
h(\bm{x}) \in (-M_2,0)\cup (0,-M_1) := \mathcal{H}.
\end{equation*}
We design a sequential proposal $g(\bm{X})$ that is more likely to 
generate $\bm{X}$ with $h(\bm{X}) \in \mathcal{H}$. 
Suppose that we have generated $X_1,\cdots,X_{k-1}\;(1\leq k \leq w)$ from this proposal.
Let 
$h_{k-1}=\tilde{\beta}_0+\sum_{i=1}^{k-1}\tilde{\beta}_{iX_i}$,
in particular $h_0=\tilde{\beta}_0$.
We determine $B^{(L)}_{k+1}=\sum_{i>k} \min_j \tilde{\beta}_{ij}$ and $B^{(U)}_{k+1}=\sum_{i>k} \max_j \tilde{\beta}_{ij}$,
the bounds for $\sum_{i>k} \tilde{\beta}_{iX_i}$.
If $X_k=j$ then the range for $h(\bm{X})$ is
\begin{equation*}
h_{k-1}+\tilde{\beta}_{ij}+[B^{(L)}_{k+1},B^{(U)}_{k+1}]:=[L_{kj},U_{kj}]. 
\end{equation*}
The larger the overlap between 
this interval and $\mathcal{H}$, the more likely that $\bm{X}$ will belong to the desired
set $\mathcal{B}^f(\hat{\bm{\beta}}^m)$. Thus, we propose $X_k$ with probability 
\begin{equation} \label{eq:seqpropose}
g_k(X_k=j \mid X_1,\cdots,X_{k-1}) \propto \left| [L_{kj}-\epsilon,U_{kj}+\epsilon]\cap \mathcal{H} \right|,
\end{equation}
where $|\,\cdot\,|$ returns the length of an interval and
$\epsilon$ is a small positive number to allow the generation of $X_k=j$ when $L_{kj}=U_{kj} \in \mathcal{H}$.
Proposing $X_k$ sequentially by (\ref{eq:seqpropose}) for $k=1,\cdots,w$ generates an $\bm{X}$ from $g(\bm{X})$. 
With $N$ proposed samples $\{\bm{X}^{(t)}\}_{t=1}^N$ we estimate the summation (\ref{eq:EDeltaRfm}) by
\begin{equation*}
\frac{1}{N} \sum_{t=1}^N \tanh \left|h(\bm{X}^{(t)})/2  \right| 
	\frac{\bm{\theta}_0(\bm{X}^{(t)})\bm{1}\{\bm{X}^{(t)} \in \mathcal{B}^f(\hat{\bm{\beta}}^m)\}}{g(\bm{X}^{(t)})}.
\end{equation*}
In this work, we propose $N=5\times 10^6$ samples for this importance sampling estimation.
We verified that the estimations were very close to the exact summations.
With different bounds for $h_1(\bm{x})$ and $h(\bm{x})$, this approach is applied to other similar
summations in (\ref{eq:EDeltaRmfMC}) and (\ref{eq:WMErrinFEMMC}).

\section{Numerical study} \label{sec:numericalWMM}

\begin{figure} [t]
 \includegraphics[width=0.75\linewidth]{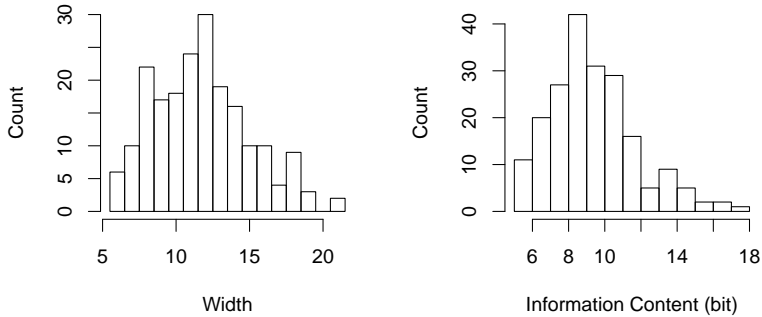}
  \caption[]{Histograms of the width and the information content of 200 WMs.}
   \label{fig:histPWM}
\end{figure}

A numerical study was performed under the WMM to confirm and quantify the lower predictive efficiency of the FE-based
estimator $\hat{\bm{\beta}}^f$ compared to the WM-based estimator $\hat{\bm{\beta}}^m$ 
discussed in Section~\ref{sec:errorWMM}. We randomly selected
200 TFs from the database TRANSFAC (Matys et al. 2003). 
For each TF, experimentally verified binding sites were used to construct a weight matrix
with a small amount of pseudo counts.
Then we randomly sampled 5,000 human upstream sequences,
each of length 10 kilo bases, and calculated their nucleotide frequency 
$\hat{\bm{\theta}}_0=$ (0.263, 0.234, 0.237, 0.266).
The 200 weight matrices display large
variability. The width $w$ ranges from 6 to 21 and the information content,
$\sum_{i=1}^w \{2+\mathbb{E}_{\bm{\theta}_i} (\log_2\theta_{iX_i})\}$,
ranges from 5.1 to 17.5 bits (Figure \ref{fig:histPWM}). 
These statistics show that our selection has covered the
typical width and strength of DNA motifs.

A constructed weight matrix was regarded as the parameter $\bm{\Theta}$
and the nucleotide frequency $\hat{\bm{\theta}}_0$ was used for the i.i.d. background in the WMM.
Since the prior odds ratio ($q_1/q_0$) of a binding site over a background site
is usually small, we chose three typical values for the inverse of
the prior odds, $\lambda=q_0/q_1=200,500,1000$, for numerical calculations.
We evaluated the AREs of the FE-based prediction w.r.t. the WM-based prediction,
defined by 
$\mbox{ARE}^m(\hat{\bm{\beta}}^f,\hat{\bm{\beta}}^m)=\alpha^m(\hat{\bm{\beta}}^f)/\alpha^m(\hat{\bm{\beta}}^m)$
in Section~\ref{sec:errorWMM}, for the 200 WMs.
As discussed in Section~\ref{sec:computation}, our evaluation of AREs was exact
for WMs of $w\leq 12$ and was carried out with simulated annealing for $w>12$.
In addition, Monte Carlo average was utilized, before simulated annealing, to approximate 
$\mbox{Cov}^m ( \sqrt{n} d\hat{\bm{\beta}}^{f} )$ (\ref{eq:covbetafwmm}) 
by simulating $5\times 10^6$ $w$-mers from the i.i.d. background.

The asymptotic relative efficiencies $\mbox{ARE}^m(\hat{\bm{\beta}}^f,\hat{\bm{\beta}}^m)$ on the 200 TFs 
are summarized in Table~\ref{tab:AREinWMM} for the three inverse prior odds. 
It is seen that for all the WMs the FE-based prediction shows lower efficiency than the WM-based prediction,
and that the median AREs of $\hat{\bm{\beta}}^f$ to $\hat{\bm{\beta}}^m$
are between 50\% and 60\% and the third quartiles ($Q_3$) between 60\% and 70\%.
Thus, for more than 75\% of the TFs, the FE procedure
is less than 70\% as efficient as the WM procedure in terms of prediction.
This confirms the loss of efficiency of the FE-based prediction under the WMM,
although both estimators are consistent. 
We note that the increase of ARE with higher $\lambda$ (smaller $q_1$)
is consistent with the lower bound defined in Proposition~\ref{lm:lossinfo}.

\begin{table}[t]  
\centering
  \caption{Summary of  $\mbox{ARE}^m(\hat{\bm{\beta}}^f,\hat{\bm{\beta}}^m)$ \label{tab:AREinWMM}}
  \vspace{0.2cm}
   \begin{tabular}{cccccc}
   \hline
   $\lambda$ & Min. & $Q_1$ & Median & $Q_3$ & Max. \\
   \hline
   200 & 0.134 & 0.391 &  0.490 &   0.595 &  0.849 \\
   500 & 0.183 &  0.475 &  0.555 &   0.638 &  0.849 \\
   1000 & 0.217 &  0.508 &  0.560 &  0.675 &  0.918 \\ 
    \hline 
  \end{tabular}
  
  \vspace{0.2cm}$Q_{1,3}$: the first and the third quartiles.
   \end{table}

\section{Applications}\label{sec:applications}

In this section, we apply the WM and the FE approaches to ChIP-seq data and 
protein binding microarray (PBM) data. We perform cross validation (CV) with training
data of different size, ranging from 20 to 500 binding sites, for two purposes.
First, with the large scale of both
types of data, we can compare empirical error rates in cross validation against theoretical error rates. 
This may allow us to verify some of the model assumptions and propose further improvement on the models.
Second, we are also interested in examining the practical performance of the two
computational methods when the number of observed binding sites varies in a wide range, 
which will provide useful guidance for future applications.

\subsection{ChIP-seq data}

In the recent two years, the ChIP-seq technique (Johnson et al., 2007; Mikkelsen et al., 2007; Robertson et al., 2007) 
has become a powerful high-throughput method to detect 
TFBS's in whole genome scale. 
A binding peak in ChIP-seq data can usually narrow down the location of a TFBS
to a neighborhood of 50 to 200 bps (Johnson et al., 2007).
ChIP-seq data that contain thousands of binding sites for a number of TFs have been generated in a study on
mouse embryonic stem cells (Chen et al., 2008). 
We chose five TFs, Esrrb, Oct4, STAT3, Sox2 and cMyc, in this study to compare the WM and the FE methods. 
The five TFs all have well-defined weight matrices
in literature and each contains more than 2,000 detected binding peaks in ChIP-seq, 
and their data quality was confirmed by motif enrichment analysis in Chen et al. (2008).
To identify the exact binding site of a ChIP-seq binding peak, we searched the 200-bp neighborhood of the peak,
100 bps on each side, to find the best match to the known weight matrix of the TF. 
Given the very small search space, the uncertainty in the exact location of the binding site should be minimal.
If the motif width of a TF is $w$, background $w$-mers were extracted from genomic control regions 
that match the locations of the binding sites relative to nearby genes.
The ratio of the number of background sites over the number of binding sites was set to 200 for every TF, that is,
the inverse prior odds ratio $\lambda=q_0/q_1=200$.
A transition matrix was estimated from the extracted background sites for each TF,
since the log Bayes factor of a Markov background model over an i.i.d model was $> 10^5$.

Based on the way we composed the data sets, the WMM with Markov background (Section~\ref{sec:WMMwithMCbg}) 
seems a more plausible data generation model. Clearly, a data set
was a mixture of detected binding sites and random background sites, 
and the background distribution was close to a Markov chain. If there is no within-motif dependence,
binding sites can be regarded as being generated from a WM model, and consequently,
the WM-based prediction is expected to have a smaller error rate compared to the FE-based prediction. 
However, if there exists within-motif dependence in binding sites, the FEM, which is able to capture such dependence, may outperform
the WM approach regardless of the mixture nature of the data sets. We computed theoretical
error rates of the two approaches under the WMM with Markov background.
For each TF, we estimated a WM from all the binding sites and 
a transition matrix from the background sites. Regarding them as the model parameters, we calculated the asymptotic
error rate of the WM-based prediction, which is the ideal error rate (\ref{eq:IdErrWMMMC}), 
and the incremental rate of the FE-based prediction (\ref{eq:EDeltaRmfMC}). 
Note that the bias due to mis-specification
of the constant term in the FE approach needs to be included for the calculation of equation (\ref{eq:EDeltaRmfMC}). 
These theoretical error rates are reported in Table~\ref{tab:ChIPseq} (the column of $n^+ = \infty$).

To compare with theoretical results, we performed cross validation to compute empirical error rates of
the WM and the FE procedures on each data set. We randomly sampled (without 
replacement) $n^+$ binding sites and $\lambda \cdot n^+$ background sites from a full data set to form a 
training set. Both approaches were applied to the training set to estimate their respective decision functions.
For WM-based prediction, a WM and a transition matrix were estimated from the training data set to construct
a decision function (\ref{eq:decisionWMMMC}) with $\beta_0=-\log(\lambda)$.
For FE-based prediction, we applied logistic regression
to the training set to obtain $\hat{h}^f(\bm{x})=\hat{\beta}_0+\bm{x} \hat{\bm{\beta}}^f$.
Then we predicted the class labels of the remaining unused sequences (test set) by each of the two decision functions and calculated
empirical error rates (CV error rates). This procedure was repeated 100 times independently for each value of $n^+$
to obtain the average CV error rate. To examine performance with a varying sample size (the number of sequences
in a training set), we chose $n^+$ from 20 to 500.

\begin{table}[t]  
  \caption{Predictive error rates (in the unit of $10^{-3}$) for ChIP-seq data \label{tab:ChIPseq}}
  \vspace{0.2cm}
   \begin{tabular}{c|ccccccc} 
   \hline
TF			& $n^+$&   20 	& 50 & 100 & 200	& 500  & $\infty$ \\
   \hline
   	          	&	WM & 2.66	& 2.49	& 2.43	& 2.40 & 2.36 & 2.30 \\
Esrrb		& FE & 4.30	& 2.77	& 2.54	& 2.45 & 2.37 & 2.45 \\
			&   (FE-WM)/WM (\%) 	& 61.7	& 11.2& 4.5 & 2.1 & 0.4 & 6.5 \\
\hline
   	          	&	WM &  3.68	& 3.54	& 3.50	& 3.47& 3.44 & 2.98 \\
Oct4		&	FE 	&  4.81	& 3.71	& 3.53	& 3.45 & 3.41 & 3.06\\
			&   (FE-WM)/WM (\%) 	&  30.7	& 4.8 & 0.9 & $-0.6$ & $-0.9$ & 2.7 \\
\hline
   	          	&	WM&   3.03 & 2.84	& 2.78	& 2.72 & 2.69 & 2.57\\
STAT3		&	FE 	&   4.69	& 3.09	& 2.84	& 2.75 & 2.70 & 2.74\\
			&   (FE-WM)/WM (\%) 	&54.8 & 8.8 & 2.2 & 1.1 & 0.3 & 6.6\\
\hline
   	          	&	WM&   2.95	& 2.75	& 2.68	& 2.65 & 2.63 & 2.53 \\
Sox2		&	FE &   3.44	& 2.89	& 2.73	& 2.68 & 2.66 & 2.59 \\
			&   (FE-WM)/WM (\%) 	&  16.6 	& 5.1 & 1.9 & 1.1 & 1.1 & 2.4 \\
\hline
   	          	&	WM &   2.67	& 2.49	& 2.42	& 2.38 & 2.34 & 2.07 \\
cMyc		&	FE 	&   3.30 & 2.51	& 2.34	& 2.26 & 2.23 & 2.24\\
			&   (FE-WM)/WM (\%) 	&   23.6	& 0.8 & $-3.3$ & $-5.0$ & $-4.7$ & $8.2$\\
   \hline
  \end{tabular}
\begin{center} 
Note: Reported are average error rates over 100 CVs. 
 \end{center}
  \end{table}
  
The average CV error rates are reported in Table~\ref{tab:ChIPseq}. The theoretical results
give a reasonable approximation to the CV error rates for both approaches when the training sample size $n^+ \geq 200$.
The asymptotic error rates of the WM approach are uniformly lower than its CV error rates for all
the TFs, while the FE approach achieves a smaller CV error rate with $n^+=500$ than its asymptotic rate for three TFs.
Consequently, the incremental percentage of the FE-based prediction  for $n^+=500$ is less than the expected level calculated
from the theory. This comparison implies that the WMM may not match the exact underlying
data generation process, although it is more plausible than the FEM given the mixture composition of the data sets. As we
discussed, potential dependence within a motif may cause possible violation to the WMM.
To verify our hypothesis, we conducted the $\chi^2$-test for every pair of motif positions 
($X_i$ and $X_k$, $1\leq i <k \leq w$) given the binding sites in
each data set. At the significance level of 0.005, we identified 25, 19, 17, 8, and 12 pairwise correlations
for Esrrb, Oct4, STAT3, Sox2, and cMyc binding sites, respectively, which gives a 
false discovery rate of $< 2\%$ for all the TFs. By capturing such correlations the FEM is
able to achieve comparable or even slightly better prediction than the WMM
with a moderate-size training sample ($n^+ \geq 100$, Table~\ref{tab:ChIPseq}). 
Finally, it is important to note that even under the exact model assumptions of the WMM, the FE-based prediction
only results in a marginal increment in error rate ($<10\%$) compared to the WM approach asymptotically
(Table~\ref{tab:ChIPseq}, $n^+=\infty$).
Together with the superior or comparable CV performance when the training size is reasonably large, this result suggests
the use of the FE approach, 
when we have a sufficient number of observed binding sites. 

\subsection{PBM data}

Protein binding microarrays (Mukherjee et al., 2004) provide a high throughput means to interrogate 
protein binding specificity to DNA sequences. Quantitative measurement
of the binding specificity of a protein to every short nucleotide sequence designed on a DNA microarray can be
obtained simultaneously.
The PBM data in Berger et al. (2008) quantified DNA binding of homeodomain proteins
via the calculation of an enrichment score, with an expected false discovery rate (FDR), 
for each double-stranded nucleotide sequence of length eight ($w=8$). 
The data set for each protein contains 32,896 8-mers, each with an enrichment score and an FDR.
We identified as the consensus binding pattern for a protein the 8-mer with the highest enrichment score, and then
labeled as binding sites those 8-mers whose FDR $<0.005$ and which differ by no more than three nucleotides
from the consensus after considering both the forward and the reverse complement strands.
The remaining 8-mers were labeled as background sites and we randomly determined their strands (orientations) to
avoid potential artifacts.
In this study we included five proteins, Hoxa11, Irx3, Lhx3, Nkx2.5, and Pou2f2, each from a different family, and
called 134, 190, 267, 145, and 213 binding sites, respectively.

The FEM, developed by the biophysics of protein-DNA binding, is expected
to be a better model that matches the design of PBM data than the WMM. 
Thus,  theoretical analysis was conducted under the FEM for the five PBM data sets.
We applied logistic regression to estimate  $\tilde{\bm{\beta}}$ and $\tilde{\beta}_0$ (\ref{eq:logistic}) with
all the labeled 8-mers in a data set, where the 8-mer `AAAAAAAA' was regarded as the reference sequence,
i.e., $\beta_{i1}\equiv 0$.
We calculated the ideal error rate $(R^f)^*$ (\ref{eq:FEidealrate}) of the FE-based prediction,
with an i.i.d. uniform background (by design the background distribution is uniform).
For the WM approach, we chose $\hat{\beta}_0$ as the log-ratio of the number of
binding sites over that of background sites, and calculated its asymptotic error rate by equation (\ref{eq:EDeltaRfm}), 
in which the bias in the constant term ($\Delta \hat{\beta}_0$) was included.
The theoretical error rates are reported in Table~\ref{tab:PBM} ($n^+=\infty$), where we find that the WM approach
gives a significantly higher error rate, between 14\% and 56\%, than the FE approach.

 \begin{table}[t]  
\begin{center}
  \caption{Predictive error rates (in the unit of $10^{-3}$) for PBM data \label{tab:PBM}}
  \vspace{0.2cm}
   \begin{tabular}{c|ccccc} 
   \hline
Protein			& $n^+$&   20 	& 50 	& 100	 & $\infty$ \\
   \hline
   	          	&	FE	& 3.57	& 2.62	& 2.43	& 1.63 \\
Hoxa11	&	WM & 3.51	& 3.22	& 3.05	& 2.10 \\
			&   (WM-FE)/FE (\%) 	& $-1.7$ & 22.9 &  25.5 &  28.8 \\
\hline
   	          	&	FE	& 5.91	& 4.71	& 4.54	& 3.26  \\
Irx3		&	WM & 5.06	& 4.85	& 4.79	& 3.71 \\
			&   (WM-FE)/FE (\%) 	& $-14.4$ & 3.0 & 5.5 & 13.8 \\
\hline
   	          	&	FE	& 7.51	& 4.64	& 4.20	& 3.18 \\
Lhx3		&	WM & 6.90	& 6.54	& 6.47	& 4.97  \\
			&   (WM-FE)/FE (\%) 	& $-8.1$ & 40.9 & 54.0 & 56.3 \\
\hline
   	          	&	FE &   3.73	& 2.31	& 2.12	& 1.80\\
Nkx2.5		&	WM &   3.64	& 3.29	& 3.16	& 2.36 \\
			&   (WM-FE)/FE (\%) 	&  $-2.4$ & 42.4 & 49.1 & 31.1 \\
\hline
   	          	&	FE &   6.25	& 4.91	& 4.56 & 3.31\\
Pou2f2		&	WM &   5.79	& 5.57	&5.49	& 4.02 \\
			&   (WM-FE)/FE (\%) 	&   $-7.3$	& 13.4 & 20.4 & 21.5\\
   \hline
  \end{tabular}
   \end{center}
   \end{table}
 
The same CV procedure as in the previous section was performed on the PBM data sets to 
compare the empirical predictive error rates of the two approaches, 
with $n^+$ varying between 20 and 100 (Table~\ref{tab:PBM}).
There is a clear decreasing trend in error rate for both approaches with the increase of the training
sample size $n^+$, although for some data sets the difference between the CV error rate for $n^+=100$
and the asymptotic rate is still quite obvious. Such discrepancy is probably due to the following two reasons.
First, the parameters $(\tilde{\bm{\beta}},\tilde{\beta}_0)$ used for the calculation of asymptotic rates
were estimated from data sets which only contain 100 to 200 binding sites. This
resulted in a high variance in the estimated parameters: The median ratio of the standard error 
over the absolute value of an estimated coefficient was between 10\% and 30\% for the five data sets.
Second, the training sample size, $n^+=100$,
is still too small to achieve a comparable error rate as $n^+ \to \infty$.
However, we have already seen substantially increased error rates of the WM-based predictions 
compared to the FE-based predictions for $n^+=100$, which is very consistent with the theoretical results.
This comparison confirms that unless the training sample size is really small, using the WM approach
may degrade predictive performance dramatically if the data generation mechanism is close to the FEM.

\section{Discussion} \label{sec:discuss}

Combining results on the ChIP-seq data and the PBM data,
this study provides some general guidance for practical applications of the WM and the FE approaches,
irrespective of  underlying data generation.
When the training sample size is small, the WM procedure seems to produce fewer errors than
the FE procedure. But when we have observed enough binding sites,
the advantage of the FE procedure is clearly seen. On one hand, it gives a comparable or slightly better
prediction than the WM approach even if the WMM is more likely for the data (Table~\ref{tab:ChIPseq}, $n^+\geq 100$). 
On the other hand,
when the data are generated in a way that matches the biophysical process of protein-DNA binding such as
the PBM data, the reduction in error rate of the FE approach can be substantial compared to the WM approach
(Table~\ref{tab:PBM}, $n^+\geq 50$).
The relative performance between the two approaches reflects a typical variance-bias tradeoff.
Estimation under the WMM is simple and more robust, which typically has a smaller variance than the FEM.
For a small sample size, predictive errors are mostly caused by variance in estimation and thus,
WM-based predictions may outperform FE-based predictions. When the sample size increases,
estimation variance decreases for both approaches and the potential bias in the WM approach becomes
the main factor for predictive errors.
Given that its primary principle comes from the biophysics of protein-DNA interactions, 
the FEM has become more attractive, based on which 
many computational methods have been developed for predicting
TF-DNA binding. In these methods a weight matrix is
sometimes used as a first order approximation for computing free energy-based binding affinity.
This work suggests that this approximation must be applied with caution.
The results on the PBM data have demonstrated that 
the WM procedure may give a prediction with 50\% or more errors
compared to the FE-based decision for a reasonably large sample size (Table~\ref{tab:PBM}).

In recent years, a substantial amount of large-scale TF-DNA binding data have been generated for 
many important biological processes. 
As demonstrated by the applications to ChIP-seq data and PBM data, 
large-sample theory is able to provide valuable insights on statistical estimation
and prediction for such large-scale data. The results in this article 
can be regarded as a first step towards a theoretical development on computational approaches
for gene regulation analysis. 
Incorporation of within-motif dependence in the WMM and interaction effects in the FEM is 
a direct next step of this work, for which the model selection component needs to be considered
in a theoretical analysis. Although desired,
further generalizations to methods for
{\it de novo} motif discovery, identification of {\it cis}-regulatory modules 
and predictive modeling of gene regulation will be more challenging future directions.

\section*{Appendices}

\subsection*{Appendix A: Proof of Proposition~\ref{lm:lossinfo}}

\begin{proof}

Let $\theta_{k(-j)} = 1-\theta_{kj}$ for $k =0,i$. 
The second order partial derivative of the marginal log-likelihood  
$l(\bm{\Theta} \mid \bm{X})  =   \log \{ q_0 \bm{\theta}_0(\bm{X})+ q_1 \bm{\Theta}(\bm{X}) \}$
w.r.t. $\theta_{ij}$ is
\begin{equation*}
\frac{\partial^2 l(\bm{\Theta} \mid \bm{X} )}{\partial \theta_{ij}^2} 
 =  - \frac{q_1^2 \{ \bm{\Theta}_{[-i]}(\bm{X}_{[-i]}) \}^2 }{ \{ q_0 \bm{\theta}_0(\bm{X})+ q_1 \bm{\Theta}(\bm{X})\}^2},
\end{equation*}
where $\bm{\Theta}(\bm{X}) =  \bm{\Theta}_{[-i]}(\bm{X}_{[-i]}) \cdot \theta_{iX_i}$ for $X_i=j,(-j)$
and similarly for $\bm{\theta}_0(\bm{X})$.
Thus, the Fisher information on $\theta_{ij}$ given $\bm{X}$ is 
\begin{eqnarray*}
I(\theta_{ij} \mid \bm{X}) & = &  - \mathbb{E} \left\{ \frac{\partial^2 l(\bm{\Theta} \mid \bm{X} )}{\partial \theta_{ij}^2}\right\}
= \sum_{\bm{x} } \frac{q_1^2 \{\bm{\Theta}_{[-i]}(\bm{x}_{[-i]})\}^2 }{ q_0 \bm{\theta}_0(\bm{x})+ q_1 \bm{\Theta}(\bm{x})} \\
& =& q_1 \sum_{\bm{x}}  \frac{q_1 \bm{\Theta}(\bm{x})}{q_0 \bm{\theta}_0(\bm{x})+ q_1 \bm{\Theta}(\bm{x})}
		\cdot \frac{1}{\theta_{ix_i}} \cdot \bm{\Theta}_{[-i]}(\bm{x}_{[-i]}) \\
& = & q_1\sum_{x \in \{j, (-j)\}} \frac{1}{\theta_{ix}} \cdot \mathbb{E}_{\bm{\Theta}_{[-i]}} \left\{  \left(  
\frac{q_0 \theta_{0x}\bm{\theta}_0(\bm{X}_{[-i]})}{q_1 \theta_{ix}\bm{\Theta}_{[-i]}(\bm{X}_{[-i]})} + 1\right)^{-1} \right\}.
\end{eqnarray*}
Because $\mathbb{E}_{\bm{\Theta}_{[-i]}} \left\{  \bm{\theta}_0(\bm{X}_{[-i]}) / \bm{\Theta}_{[-i]}(\bm{X}_{[-i]}) \right\} =1$, 
Jensen's inequality implies that
\begin{equation*}
I(\theta_{ij} \mid \bm{X})  \geq  \sum_{x\in\{j,(-j)\}} \frac{q_1^2}{q_0 \theta_{0x} + q_1 \theta_{ix}}=\frac{q_1^2}{\bar{\theta}_{ij}(1-\bar{\theta}_{ij})}.
\end{equation*}
The lower bound $B (q_1,{\theta}_{ij},{\theta}_{0j})$ is obtained by dividing the R.H.S. of this inequality by 
the Fisher information on $\theta_{ij}$ given $\bm{X}$ and $Y$ jointly,
\begin{equation*} 
I(\theta_{ij} \mid \bm{X},Y) = - \mathbb{E} \left\{ \frac{\partial^2 l(\bm{\Theta} \mid \bm{X},Y )}{\partial \theta_{ij}^2}\right\}
= \frac{q_1}{\theta_{ij}(1-\theta_{ij})},
\end{equation*}
where $ l(\bm{\Theta} \mid \bm{X},Y )=\log P(\bm{X},Y \mid \bm{\Theta} )$ is the joint log-likelihood. 
\end{proof}

\subsection*{Appendix B: Derivation of $\mathbb{E}[ \Delta R_1^f(\hat{\bm{\beta}}^m)]$ (\ref{eq:WMErrinFEMMC})}

Given the estimated weight matrix
$\hat{\bm{\Theta}}^m$ based on observed binding sites $\bm{D}_n^+$,  
the constructed decision function of the WM approach
\begin{eqnarray}\label{eq:hath1m}
\hat{h}_1^m(\bm x)&=&\log(q_1/q_0)+\sum_{i=1}^w [\log\hat{\theta}^m_{ix_i}-\log\psi_0(x_{i-1},x_i)] \notag\\
&\toinP& \tilde{\beta}_0 + \sum_{i=1}^w \left\{ \log \frac{{\theta}^f_{ix_i}}{\theta^f_{is_i}} - 
\log \frac{\psi_0(x_{i-1}, x_{i})}{\psi_0(s_{i-1}, s_i)}  \right\}, \mbox{ as $n\rightarrow \infty$, }
\end{eqnarray}
where $\theta^f_{ij}= P(X_i=j \mid Y=1)$ under the FEM with Markov background, 
$(s_1,\cdots,s_w)$ is the reference sequence, and 
\begin{equation*}
\tilde{\beta}_0=\log(q_1/q_0)+\sum_{i=1}^w \log\{\theta^f_{is_i}/\psi_0(s_{i-1},s_i)\}.
\end{equation*}
Let $\bm{x}_{[-i]}(s)=(x_1,\cdots,x_{i-1}, s,  x_{i+1},\cdots,x_w)$. 
Following a similar derivation in Section~\ref{sec:errorFEM}, we have
$\theta^f_{ij} \propto \exp(\tilde{\beta}_{ij}+\eta_{ij})$, where
\begin{equation*}
\eta_{ij}=\log \left\{ \sum_{\bm{x}_{[-i]}} \frac{e^{u_i}}{1+e^{\tilde{\beta}_{ij}} e^{u_i} } \bm{\psi}_0(\bm{x}_{[-i]}(j)) \right\}
 -\log \left\{ \sum_{\bm{x}_{[-i]}} \frac{e^{u_i}}{1+ e^{u_i} } \bm{\psi}_0(\bm{x}_{[-i]}(s_i)) \right\}
\end{equation*}
with $u_i=\tilde{\beta}_0+\bm{x}_{[-i]}\tilde{\bm{\beta}}_{[-i]}$. 
Since $\tilde{\beta}_{is_i}=\eta_{is_i}= 0$, 
$\log(\theta^f_{ij}/\theta^f_{is_i})= \tilde{\beta}_{ij}+\eta_{ij}$ for all $i$ and $j$.
Thus, equation (\ref{eq:hath1m}) becomes
\begin{equation*} 
\hat{h}_1^m(\bm{x})  \toinP  \tilde{\beta}_0 + \sum_{i=1}^w \left\{ \tilde{\beta}_{ix_i} + \eta_{ix_i} - 
\log \frac{\psi_0(x_{i-1}, x_{i})}{\psi_0(s_{i-1}, s_i)}  \right\} 
 =   h(\bm{x}) + \delta(\bm{x}), 
\end{equation*}
where $\delta(\bm{x})=\sum_{i=1}^w \eta_{ix_i} - \log \{ \psi_0(x_{i-1}, x_{i}) / \psi_0(s_{i-1}, s_i)\}$.
Let $\Delta \hat{h}_1^m(\bm{x})= \hat{h}_1^m(\bm{x})-h(\bm{x})$.
The asymptotic normality of $\sqrt{n} d \hat{\bm{\Theta}}^m$ 
implies that $\sqrt{n}\{ \Delta \hat{h}_1^m(\bm{x}) - \delta(\bm{x}) \}$
follows a limiting normal distribution with mean 0 and a finite (possibly zero) variance for every $\bm{x}$. 
Equation (\ref{eq:WMErrinFEMMC}) then follows from Theorem~\ref{th:BiasedIncRate}.

\section*{Acknowledgements}

The author thanks Wing H. Wong, Jun S. Liu and
Zhengqing Ouyang for helpful discussions.
This work was supported by NSF grant DMS-0805491.

\end{document}